\documentclass[aps,prd,reprint,onecolumn,notitlepage,groupedaddress]{revtex4-1}
\usepackage{extraipa}
\usepackage{amsmath}
\usepackage{amsfonts}
\usepackage{amssymb}

\newcommand{\etal}{\textit{et al}.}

\begin{document}

\title{One-Loop Effective Action for Nonminimal Natural Inflation Model}

\author{Sandeep Aashish}%
\email[]{sandeepa16@iiserb.ac.in}

\author{Sukanta Panda}
\email[]{sukanta@iiserb.ac.in}

\affiliation{Department of Physics, Indian Institute of Science Education and Research, Bhopal 462066, India}

%


\date{\today}

\begin{abstract}
Recent and upcoming experimental data as well as the possibility of rich phenomenology has spiked interest in studying the quantum gravitational effects in cosmology at low (inflation-era) energy scales. While Planck scale physics is under development, it is still possible to incorporate quantum gravity effects at relatively low energies using quantum field theory in curved spacetime, which serves as a low-energy limit of planck scale physics. We use the Vilkovisky-DeWitt's covariant effective action formalism to study quantum gravitational corrections to a recently proposed Natural Inflation model with periodic nonminimal coupling. We present the 1-loop effective action for this theory valid in the flat-potential region, considering perturbative corrections upto quadratic order in background scalar fields.
\end{abstract}



\maketitle

\section{\label{intro}Introduction}
Natural inflation was first introduced by Freese \etal \cite{freese1990} as an approach where inflation arises dynamically (or \textit{naturally}) from particle physics models. In natural inflation models, a flat potential is effected using pseudo Nambu-Goldstone bosons arising from breaking the continuous shift symmetry of Nambu-Goldstone modes into a discrete shift symmetry. As a result, the inflation potential in a Natural inflation model takes the form,
\begin{eqnarray}
\label{eq01}
V(\phi) = \Lambda^4 \left(1 + \cos(\phi/f)\right)
\end{eqnarray} 
where the magnitude of parameter $\Lambda^4$ and periodicity scale $f$ are model dependent. However, majority of natural inflation models are in tension with recent Planck 2018 results \cite{planck2018x}. 

Recently, an extension of the original natural inflation model was proposed that introduces a new periodic non-minimal coupling to gravity \cite{ferreira2018}. This modification leads to a better fit with observation data, with $n_{s}$ and $r$ values well within $95\%$ C.L. region. from combined Planck 2018+BAO+BK14 data. Moreover, $f$ becomes sub-planckian, contrary to a super-planckian $f$ in the original natural inflation model, and addresses issues related to gravitational instanton corrections.

Our objective here is to study one-loop quantum gravitational corrections to the natural inflation model with non-minimal coupling, using Vilkovisky-DeWitt's covariant effective action approach \cite{parker2009}. A caveat is that effective action cannot be calculated exactly, so only perturbative results are feasible. Hence, we apply a couple of approximations. First, we work in the regime where potential is flat, i.e. $\phi \ll f$. Second, the background metric is Minkowski.  

The action for the nonminimal natural inflation in the Einstein frame is given by,
\begin{eqnarray}
\label{eq02}
S = \int d^4 x \sqrt{-g}\left(-\dfrac{2 R}{\kappa^2} + \dfrac{1}{2}K(\phi)\phi {}_{;a} \phi {}^{;a} + \dfrac{V(\phi)}{(\gamma(\phi))^4}\right)
\end{eqnarray}
where, 
\begin{equation}
\label{eq03}
\gamma(\phi)^2 = 1 + \alpha\left(1+\cos\left(\dfrac{\phi}{f}\right)\right),
\end{equation}
and,
\begin{equation}
\label{eq04}
K(\phi) = \dfrac{1 + 24\gamma'(\phi)^2/\kappa^2}{\gamma(\phi)^2}. 
\end{equation}
$V(\phi)$ is as in Eq. (\ref{eq01}). 
In the region of where potential is flat, $\phi/f \ll 1$, and we expand all periodic functions in Eq. (\ref{eq02}) upto quartic order in $\phi$:
\begin{equation}
\label{eq05}
S = \int d^4 x \sqrt{-g} \left(- \frac{2 R}{\kappa^2} + \tfrac{1}{2} m^2 \phi^2 + \tfrac{1}{24} \lambda \phi^4 + \tfrac{1}{2} (k_0 + k_1 \phi^2) \phi {}_{;a} \phi {}^{;a}\right) + \mathcal{O}(\phi^5)
\end{equation}
where parameters $m, \lambda,k_0$ and $k_1$ have been defined out of $\alpha, f$ and $\Lambda^4$ in from Eq. (\ref{eq02}):
\begin{eqnarray}
\label{param}
m^2 &=& \dfrac{\Lambda^4 (2\alpha - 1)}{(1 + 2\alpha)^3 f^2};\nonumber \\
\lambda &=& \dfrac{\Lambda^4 (8\alpha^2 - 12\alpha + 1)}{(1 + 2\alpha)^4 f^4}; \nonumber \\
k_0 &=& \dfrac{1}{1 + 2\alpha}; \nonumber \\
k_1 &=& \dfrac{\alpha(\kappa^2 f^2 + 96\alpha^2 + 48\alpha)}{2\kappa^2 f^4 (1 + 2\alpha)^2}. \nonumber \\
\end{eqnarray}
We have also omitted a constant term appearing in $\ref{eq05}$ because such terms are negligibly small in early universe.

\section{Effective action formalism}
The quantization of a theory $S[\varphi]$ with fields $\varphi^{i}$ is performed about a classical background $\bar{\varphi}^{i}$:  $\varphi^{i} = \bar{\varphi}^{i} + \zeta^{i}$, where $\zeta^{i}$ is the quantum part. In our case, $\varphi^{i}=\{g_{\mu\nu},\phi\}$; $\bar{\varphi}^{i}=\{\eta_{\mu\nu},\bar{\phi}\}$ where, $\eta_{\mu\nu}$ is the Minkowski metric; and, $\zeta^{i}=\{\kappa h_{\mu\nu},\delta\phi\}$.

The 1-loop effective action is given by
\begin{eqnarray}
\label{eq06}
\Gamma = -\ln\int[d\zeta]\exp\left[\zeta^{i}\zeta^{j}\Big(S_{,ij}[\bar{\varphi}] - \bar{\Gamma}^{k}_{ij}S_{,k}[\bar{\varphi}]\Big) + \frac{1}{2\alpha}\chi_{\beta}^{2}\right]-\ln\det Q_{\alpha\beta}[\bar{\varphi}]
\end{eqnarray}
as $\alpha\longrightarrow 0$ (Landau-DeWitt gauge). Here, $S_{,i}$ and $S_{,ij}$ are first and second functional derivative w.r.t the fields $(g_{\mu\nu},\phi)$ at background $(\eta_{\mu\nu},\bar{\phi})$, respectively. $\bar{\Gamma}^{k}_{ij}$ are Vilkovisky-DeWitt connections that ensure covariance. $\chi_{\beta}$ is the gauge condition for the GCT symmetry, and $Q_{\alpha\beta}$ is the ghost term that appears during quantization. Fortunately, upto quadratic order in background fields, and upto $\kappa^2$ order in quartic order terms, ghost determinant does not contribute. Hence, we will omit writing $Q_{\alpha\beta}$ in our calculations. 

For convenience, we write the exponential in the first term of $\Gamma$ as,
\begin{eqnarray}
\label{eq07}
\exp[\cdots] &=& \exp\left(\tilde{S}[\bar{\varphi}^{0}]+\tilde{S}[\bar{\varphi}^{1}]+\tilde{S}[\bar{\varphi}^{2}]+\tilde{S}[\bar{\varphi}^{3}]+\tilde{S}[\bar{\varphi}^{4}]\right)\nonumber \\
&\equiv & \exp(\tilde{S}_{0}+\tilde{S}_{1}+\tilde{S}_{2}+\tilde{S}_{3}+\tilde{S}_{4})
\end{eqnarray}
Treating $\tilde{S}_{1},...,\tilde{S}_{4}$ terms as perturbative, the final contribution to $\Gamma$ at each order of $\bar{\varphi}$ is:
\begin{eqnarray}
\label{eq08}
\mathcal{O}(\bar{\varphi}) &=& 0\nonumber \\
\mathcal{O}(\bar{\varphi}^2) &=& \langle\tilde{S}_{2}\rangle - \dfrac{1}{2}\langle\tilde{S}_{1}^{2}\rangle \nonumber \\
\mathcal{O}(\bar{\varphi}^3) &=& 0\nonumber \\
\mathcal{O}(\bar{\varphi}^4) &=& \langle\tilde{S}_{4}\rangle - \langle\tilde{S}_{1}\tilde{S}_{3}\rangle + \mathcal{O}(\kappa^4) - \ln\det Q_{\alpha\beta}
\end{eqnarray}
The correlators are calculated using Wick's theorem and basic propagator relations
\begin{eqnarray}
\label{eq09}
 \langle h_{\mu\nu}(x)h_{\rho\sigma}(x')\rangle &=& G_{\mu\nu\rho\sigma}(x.x'); \nonumber \\
  \quad \langle \delta\phi(x)\delta\phi(x')\rangle &=& G(x,x'); \nonumber \\
 \langle h_{\mu\nu}(x)\delta\phi(x')\rangle &=& 0,
\end{eqnarray}
where, $G(x,x')$ and $G_{\mu\nu\rho\sigma}(x,x')$ are respective green's functions for gravity and scalar field. 

\section{Results and Conclusions}
In what follows, an integration over coordinates ($\int d^4 x$) and coordinate dependence is assumed unless stated otherwise. Our results for $S_{0,1,2}$ are as follows:
\begin{eqnarray}
\label{result1}
S_0 &=& \tfrac{1}{2} m^2 (\delta \phi)^2 + \tfrac{1}{2} k_0 \delta \phi{}_{,a} \delta \phi{}^{,a} + h^{ab} h^{}{}_{a}{}^{c}{}_{,b}{}_{,c} -  \frac{h^{ab} h^{}{}_{a}{}^{c}{}_{,b}{}_{,c}}{\alpha} -  h^{a}{}_{a} h^{bc}{}_{,b}{}_{,c} \nonumber \\ && + \frac{h^{a}{}_{a} h^{bc}{}_{,b}{}_{,c}}{\alpha} -  \tfrac{1}{2} h^{ab} h^{}{}_{ab}{}^{,c}{}_{,c} + \tfrac{1}{2} h^{a}{}_{a} h^{b}{}_{b}{}^{,c}{}_{,c} -  \frac{h^{a}{}_{a} h^{b}{}_{b}{}^{,c}{}_{,c}}{4 \alpha} \\
\label{result2}
S_1 &=& \tfrac{1}{2} m^2 \kappa \delta \phi h^{a}{}_{a} \phi - 
\frac{1}{4} m^2 \kappa \nu \delta \phi h^{a}{}_{a} \phi -  
\frac{1}{2} k_0 \kappa \delta \phi h^{b}{}_{b} \phi {}^{,a}{}_{,a} 
+ \tfrac{1}{4} k_0 \kappa \nu \delta \phi h^{b}{}_{b} \phi 
{}^{,a}{}_{,a} \nonumber \\ 
&& -  \tfrac{1}{2} k_0 \kappa \delta \phi h^{b}{}_{b}{}_{,a} \phi \
{}^{,a} + \frac{\kappa \omega \delta \phi h^{b}{}_{b}{}_{,a} \
\phi {}^{,a}}{2 \alpha} + k_0 \kappa \delta \phi \phi {}^{,a} \
h^{}{}_{a}{}^{b}{}_{,b} -  \frac{\kappa \omega \delta \phi \phi \
{}^{,a} h^{}{}_{a}{}^{b}{}_{,b}}{\alpha} \nonumber \\ 
&& + k_0 \kappa \delta \phi h^{}{}_{ab} \phi {}^{,a}{}^{,b} \\
\label{result3}
S_2 &=& - \tfrac{1}{8} m^2 \kappa^2 h^{}{}_{ab} h^{ab} \phi^2 + \tfrac{1}{16} m^2 \kappa^2 h^{a}{}_{a} h^{b}{}_{b} \phi^2 + \tfrac{1}{4} \lambda \phi^2 (\delta \phi)^2 -  \tfrac{1}{8} m^2 \kappa^2 \nu \phi^2 (\delta \phi)^2 \nonumber \\ 
&& + \tfrac{1}{2} k_1 \phi^2 \delta \phi{}_{,a} \delta \phi{}^{,a} + 2 k_1 \delta \phi \phi \phi {}_{,a} \delta \phi{}^{,a} -  \tfrac{1}{8} k_0 \kappa^2 h^{}{}_{bc} h^{bc} \phi {}_{,a} \phi {}^{,a} + \tfrac{1}{16} k_0 \kappa^2 \nu h^{}{}_{bc} h^{bc} \phi {}_{,a} \phi {}^{,a} \nonumber \\ 
&& + \tfrac{1}{16} k_0 \kappa^2 h^{b}{}_{b} h^{c}{}_{c} \phi {}_{,a} \phi {}^{,a} -  \tfrac{1}{32} k_0 \kappa^2 \nu h^{b}{}_{b} h^{c}{}_{c} \phi {}_{,a} \phi {}^{,a} + \tfrac{1}{2} k_1 (\delta \phi)^2 \phi {}_{,a} \phi {}^{,a} -  \tfrac{1}{16} k_0 \kappa^2 \nu (\delta \phi)^2 \phi {}_{,a} \phi {}^{,a} \nonumber \\ 
&& + \frac{\kappa^2 \omega^2 (\delta \phi)^2 \phi {}_{,a} \phi {}^{,a}}{4 \alpha} + \tfrac{1}{2} k_0 \kappa^2 h^{}{}_{a}{}^{c} h^{}{}_{bc} \phi {}^{,a} \phi {}^{,b} -  \tfrac{1}{4} k_0 \kappa^2 \nu h^{}{}_{a}{}^{c} h^{}{}_{bc} \phi {}^{,a} \phi {}^{,b} -  \tfrac{1}{4} k_0 \kappa^2 h^{}{}_{ab} h^{c}{}_{c} \phi {}^{,a} \phi {}^{,b} \nonumber \\ 
&& + \tfrac{1}{8} k_0 \kappa^2 \nu h^{}{}_{ab} h^{c}{}_{c} \phi {}^{,a} \phi {}^{,b}
\end{eqnarray}
In our case, $S_0$ leads to the free theory propagators for gravity and massive scalar field, which are well known. $S_1$ contains cross terms of $h_{\mu\nu}$ and $\delta\phi$, and does not contribute at $\mathcal{O}(\bar{\phi})$, since $\langle S_{1}(x)\rangle = 0$. Contribution from $S_{1}$ comes at $\mathcal{O}(\bar{\phi}^2)$ from terms consisting of $\langle S_{1}(x)S_{1}(x')\rangle$ (Eq. \ref{eq08}). Expectation value of $S_{2}(x)$ describes contributions from tadpole diagrams. $\langle S_{1}(x)S_{1}(x')\rangle$  encompasses interaction terms between gravity and scalar field. For bookkeeping, parameters like $\omega$ and $\nu$ have been introduced to track gauge dependent terms and Vilkovisky-DeWitt terms respectively. Playing with these parameters is an interesting exercise to check the gauge invariant nature of Vilkovisky-DeWitt approach \cite{mackay2010}. We present here the divergent part of $Gamma$ at $\mathcal{O}(\bar{\phi}^4)$, obtained after regularizing these path integrals using dimensional regularization (a factor of $1/(n-4)$ is assumed with the limit $n\to 4$):
\begin{eqnarray}
\label{divp1}
divp(S_{2}) &=& \frac{i k_1 m^4 \phi^2}{16 k_0^2 \pi^2} -  \frac{i m^2 \lambda \phi^2}{32 k_0 \pi^2} + \frac{i m^4 \kappa^2 \nu \phi^2}{64 k_0 \pi^2} + \frac{i k_1 m^2 \phi \phi {}^{,a}{}_{,a}}{16 k_0 \pi^2} -  \frac{i m^2 \kappa^2 \nu \phi \phi {}^{,a}{}_{,a}}{128 \pi^2} + \frac{i m^2 \kappa^2 \omega^2 \phi \phi {}^{,a}{}_{,a}}{32 k_0 \pi^2 \alpha} \\
\label{divp2}
divp(S_1 S'_1) &=& - \frac{3i m^4 \kappa^2 \phi^2}{16 \pi^2} + \frac{i m^4 \alpha \kappa^2 \phi^2}{16 \pi^2} + \frac{3i m^4 \kappa^2 \nu \phi^2}{16 \pi^2} -  \frac{i m^4 \alpha \kappa^2 \nu \phi^2}{16 \pi^2} \nonumber \\ 
&& -  \frac{3i m^4 \kappa^2 \nu^2 \phi^2}{64 \pi^2} + \frac{i m^4 \alpha \kappa^2 \nu^2 \phi^2}{64 \pi^2} + \frac{3i k_0 m^2 \kappa^2 \phi \phi {}^{,a}{}_{,a}}{16 \pi^2} -  \frac{9i k_0 m^2 \kappa^2 \nu \phi \phi {}^{,a}{}_{,a}}{32 \pi^2} \nonumber \\ 
&& + \frac{3i k_0 m^2 \alpha \kappa^2 \nu \phi \phi {}^{,a}{}_{,a}}{32 \pi^2} + \frac{3i k_0 m^2 \kappa^2 \nu^2 \phi \phi {}^{,a}{}_{,a}}{32 \pi^2} -  \frac{i k_0 m^2 \alpha \kappa^2 \nu^2 \phi \phi {}^{,a}{}_{,a}}{32 \pi^2} -  \frac{i m^2 \kappa^2 \omega \phi \phi {}^{,a}{}_{,a}}{8 \pi^2} \nonumber \\ 
&& + \frac{i m^2 \kappa^2 \omega^2 \phi \phi {}^{,a}{}_{,a}}{16 k_0 \pi^2 \alpha} -  \frac{i k_0^2 \kappa^2 \nu \phi \phi {}^{,b}{}_{,b}{}^{,a}{}_{,a}}{32 \pi^2} + \frac{i k_0^2 \alpha \kappa^2 \nu \phi \phi {}^{,b}{}_{,b}{}^{,a}{}_{,a}}{32 \pi^2} -  \frac{i k_0 m^2 \alpha \kappa^2 \phi \phi {}_{,a}{}^{,a}}{16 \pi^2} \nonumber \\ 
&& + \frac{i k_0 m^2 \alpha \kappa^2 \nu \phi \phi {}_{,a}{}^{,a}}{32 \pi^2} + \frac{i m^2 \kappa^2 \omega \phi \phi {}_{,a}{}^{,a}}{16 \pi^2} -  \frac{i m^2 \kappa^2 \nu \omega \phi \phi {}_{,a}{}^{,a}}{32 \pi^2} -  \frac{i k_0^2 \alpha \kappa^2 \phi \phi {}_{,a}{}^{,b}{}_{,b}{}^{,a}}{16 \pi^2} \nonumber \\ 
&& -  \frac{i k_0^2 \alpha \kappa^2 \nu \phi \phi {}_{,a}{}^{,b}{}_{,b}{}^{,a}}{32 \pi^2} + \frac{i k_0 \kappa^2 \omega \phi \phi {}_{,a}{}^{,b}{}_{,b}{}^{,a}}{16 \pi^2} + \frac{i k_0 \kappa^2 \nu \omega \phi \phi {}_{,a}{}^{,b}{}_{,b}{}^{,a}}{32 \pi^2} + \frac{i k_0^2 \alpha \kappa^2 \phi \phi {}^{,a}{}_{,a}{}^{,b}{}_{,b}}{16 \pi^2} \nonumber \\ 
&& + \frac{i k_0^2 \kappa^2 \nu \phi \phi {}^{,a}{}_{,a}{}^{,b}{}_{,b}}{8 \pi^2} -  \frac{i k_0^2 \alpha \kappa^2 \nu \phi \phi {}^{,a}{}_{,a}{}^{,b}{}_{,b}}{16 \pi^2} -  \frac{3i k_0^2 \kappa^2 \nu^2 \phi \phi {}^{,a}{}_{,a}{}^{,b}{}_{,b}}{64 \pi^2} + \frac{i k_0^2 \alpha \kappa^2 \nu^2 \phi \phi {}^{,a}{}_{,a}{}^{,b}{}_{,b}}{64 \pi^2}
\end{eqnarray}
Note that there are gauge dependent terms in both Eqs. (\ref{divp1}) and (\ref{divp2}), two of which diverge as $\alpha\to 0$ (Landau gauge). However, when all contributions are added to evaluate $\Gamma$, these terms vanish so that the final result is gauge-invariant. Final result for divergent part of $\Gamma$ obtained from Eq. (\ref{eq08}) after removing bookkeeping parameters ($\omega\to 1$, $\nu \to 1$) in the Landau gauge ($\alpha \to 0$): 
\begin{eqnarray}
\label{final}
divp(\Gamma) = L\int d^4 x \Big[\frac{i k_1 m^4 \phi^2}{2 k_0^2} + \tfrac{3}{16}i m^4 \kappa^2 \phi^2 + \frac{i m^4 \kappa^2 \phi^2}{8 k_0} -  \frac{i m^2 \lambda \phi^2}{4 k_0} + \frac{i k_1 m^2 \phi \phi {}^{,a}{}_{,a}}{2 k_0} + \tfrac{5}{16}i m^2 \kappa^2 \phi \phi {}^{,a}{}_{,a} \nonumber \\ -  \tfrac{3}{8}i k_0 \kappa^2 \phi \phi {}^{,a}{}_{,a}{}^{,b}{}_{,b} -  \tfrac{3}{16}i k_0^2 \kappa^2 \phi \phi {}^{,a}{}_{,a}{}^{,b}{}_{,b}\Big]
\end{eqnarray}
where $L = \dfrac{1}{8\pi^2 (n-4)}$. 

One can in principle construct counterterms from the classical action functional to absorb the divergent part in Eq. (\ref{final}), which will inturn induce 1-loop corrections to parameters of the theory (\ref{eq02}). Unfortunately, no counterterms can be found for the last two terms in Eq. (\ref{final}) which are indeed fourth order derivative terms, and highlight the issue of renormalizability of gravity theories. A subsequent work will contain more detailed calculations upto quartic order in background fields, and the effect of quantum corrections in the context of inflation \cite{aashish2019d}.
 
\begin{acknowledgments}
This work was partially funded by DST (Govt. of India), Grant No. SERB/PHY/2017041. Calculations were performed using xAct packages for Mathematica.
\end{acknowledgments}


\bibliography{refdae}

\end{document}